# Boscovich and the Brera Observatory[1]

E. Antonello

*INAF – Osservatorio Astronomico di Brera,*
*Via E. Bianchi 46, I-23807 Merate, Italy*
elio.antonello@brera.inaf.it

**Abstract.** In the mid 18th century both theoretical and practical astronomy were cultivated in Milan by Barnabites and Jesuits. In 1763 Boscovich was appointed to the chair of mathematics of the University of Pavia in the Duchy of Milan, and the following year he designed an observatory for the Jesuit Collegium of Brera. The Specola was built in 1765 and it became quickly one of the main European observatories. We discuss the relation between Boscovich and Brera in the framework of a short biography. An account is given of the initial research activity in the Specola, of the departure of Boscovich from Milan in 1773 and his coming back just before his death.

**1. Astronomy in Milan**

The Brera Observatory in Milan is one of the many European observatories built by Jesuits (Udìas 2003), and its basic characteristics, that were maintained for many years, were essentially those planned by Boscovich. Several publications have been dedicated to the relation between Boscovich and Brera, and two meetings were held here, the first in 1962 on occasion of the 250 years of his birth (AA.VV. 1963), and the second in 1987 for the 200 years of his death (AA.VV. 1988). The latter in particular marked the beginning of a renewed activity in the observatory on the research fields representing the challenges of contemporary astronomy and astrophysics, an activity that ideally joined with the fervor of Boscovich's times. In the mid 18th century the Duchy of Milan was ruled by the Austrian Habsburg house, and important reforms were introduced that allowed progresses in many fields, economy, government, education, science, arts and culture. Probably it is not by chance that even astronomy took advantage of this favorable period.

The Barnabite Paolo Frisi was interested in physics and theoretical astronomy (and also mathematics, hydraulics and engineering). While he was teaching astronomy at the Barnabite school, some Jesuits were making astronomical observations with amateur instrumentation as complement to lessons in their Collegium of Brera. Owing to the strong interest in this matter, it was decided to move an expert astronomer, the Father Louis Lagrange, from the Jesuit observatory of Marseille to Brera. Lagrange arrived in Milan at the end of 1762, and here he planned rigorously the astronomical and meteorological research. That could be considered the beginning of the professional observational astronomy in Milan. The following year, Boscovich was appointed to the chair of mathematics of the University of Pavia (the university of the Duchy). He was a polymath and gave contributions to mathematics, geometry, optics, astronomy, geodesy, engineering, hydraulics, and also to poetry; he was an architectural advisor to Popes, theologian and a cosmopolitan diplomat. His very wide interests show that the clearcut division of the cultural disciplines did not exist yet in 18th century, but with the increasing specialization of the scientific research this separation would have been inevitable. His most famous work, the *Theoria Philosophiae Naturalis* (1758, 1763), had an enormous impact on the scientific and philosophical thought of the 19h century. According to Barrow (2007), Boscovich was the first to have a scientific vision of a Theory of Everything, and the influence of his *Theoria* was wide and deep especially in Britain, where Faraday, Maxwell and Kelvin would record their indebtedness to its inspiration. The

---

philosopher Nietzsche compared Boscovich to Copernicus for his revolutionary hypothesis on the structure of matter (Nietzsche 1886).

**2. Before the year 1764**

Boscovich born 300 years ago in Ragusa (Dubrovnik, Croatia), a small republic in Dalmatia. Ragusa was struggling to keep its independence against the great powers of that time (e.g. the Ottoman Empire), and Boscovich himself will serve his country in his diplomatic activity. He got education at the primary school of Jesuits (Collegium Ragusinum), and being a well promising student he was sent off to Rome at age 14. He completed the scientific and humanistic studies in the Collegium Romanum, where he was appointed to the chair of mathematics in 1740, and after the theological studies took his vows as a priest in 1744. The years from 1744 to 1758 represent his most prolific period of mature scholarship[2]. At the same time these are also years of development as a polished Jesuit in society. After 1758, that experience gradually removed him from the Collegium Romanum into a life which was regarded as being more necessary for the Jesuit Order, then passing through difficult days, which Boscovich was committed to serve implicitly and loyally to the end (Hill 1961). He was introduced into the Accademia degli Arcadi, and this gave him the opportunity to meet an exceptionally wide circle of influential european personages.

His first mission outside the Papal States occurred in 1756, and he went as an expert hydrographer to Lucca to arbitrate a dispute that had arisen between the Republic of Lucca and the Grand Duchy of Tuscany over adjacent waters. The Tuscany was under the Austrian rule, and Boscovich went also to Vienna, where he had the opportunity to publish his first edition of the *Theoria*. When he was back in Rome in 1758, he already knew that he was to go abroad for a longer diplomatic mission, that would have included France, Germany and the Netherlands. No document has been found as yet about the political purpose and the details of such a mission. From 1756 the Seven Years' War was shacking Europe, with the involvement of Austria, France, Sweden, Russia, Spain, some Germany states, Great Britain, Prussia, Portugal and several colonies[3].

The political situation of Jesuits was increasingly difficult, and in 1758 they were expelled from Portugal. Up to that time Boscovich apparently did not resented such difficulties, since he was considered an important European scholar and was appreciated in cultural and political circles. In 1759 he was in France and visited several Jesuit Collegii, in particular Lyon and Marseille with their observatories. In Paris he had high level meetings with scholars and politicians. He had similar meetings also in Great Britain, where he visited the Greenwich observatory. The subsequent steps of the journey included the Netherlands, Germany and Lorraine in 1761, then Venice and Constantinople. During his life Boscovich suffered sometimes from disease, but when he was in Constantinople he fell ill for some months with a dangerous infection in a leg. He then recovered, but the problems with the leg will be never completely solved. In 1762 Boscovich left from Turkey and travelled to Poland in two months; he wrote a diary of this long journey, which presents great interest for it is one of the few travel records of the eighteenth century that cover that area of Turkey, Bulgaria and Moldavia. In 1763 he was in Italy, and during the summer he published the second edition of the *Theoria* in Venice, before going back to Rome. As summarized by Hill (1961), Boscovich was now in his fifty-second year. He had seen Europe from one end to the other, he had made hosts of friends and some enemies, he was versed in the trend of political events, he knew the situation of the Jesuit Order in many countries.

**3. In the Duchy of Milan**

In 1763 Boscovich received the invitation to occupy the chair of mathematics in the University of Pavia. At that time, the question of reforming Pavia University was a much discussed topic in the

---

[2] A catalogue of his works can be found in the website of the Edizione Nazionale R.G. Boscovich.
[3] For example, the North America, where the war between French and British began in 1754.

Duchy of Milan and in Vienna. The austrian Chancellor (prime minister) Kaunitz decided to enlarge the structure, increase the library, attract good professors. It is therefore not surprising that Boscovich should have been invited. Before leaving from Rome, Boscovich had to tackle the centuries old problem of the drainage of the Pontine marshes, and in the spring of 1764 he handed over to the Papal Government his study. Then he took up his Pavia appointment (Hill 1961). On april Boscovich arrived at the Brera Collegium. Probably it was during his short stay in Milan, before going to Pavia, that he was informed about the local researches in astronomy and the desire for the realisation of a Specola. He probably expressed his ideas about an observatory that had to be at the level of the best european institutes, and wrote to the Father General in Rome informing him about the Specola (Proverbio 1997). During the summer vacations, with much enthusiasm he gave his expert advice at every stage of the project. The plan of the building was entrusted to him, in view of his skill as a mathematician and structural engineer combined with his knowledge of astronomy. His remarkable plans for a modern observatory, including those for the reinforcement of the existing structure on which it would be based, were passed by the governor and by the local minister plenipotentiary Count Firmian (Hill 1961). Boscovich devoted much energy to this enterprise which was being realised at the expense of the Jesuit College and by contributions from individual Jesuits. Boscovich himself spent much of his own money on the building of the observatory and on the instrumentation. A large wooden model was produced probably at the beginning of 1765. The Father General gave his approval and the construction of the Specola began around april of that year (Proverbio 1997). The observatory was located in the south-east corner of the Brera Palace. During the summer, Boscovich went with the French astronomer Lalande in his *Voyage en Italie* (Lalande 1769). Lalande was a friend of him and of several Italian astronomers; he will give a strong help to Boscovich after the Jesuit suppression, and later on he will take care of the troubled astronomers during the Napoleonic campaign of Italy (Antonello 2010). The observatory building was completed before the end of 1765. Given the significant expenses for the construction, it was decided to purchase the main astronomical instrument, a large mural quadrant, in France, since the French instruments (Canivet) were less expensive though probably less accurate than the English ones (Bird). Also an order for a large sextant was placed to Canivet. The quadrant and the sextant are the oldest instruments still existing.

It may be worth to remark those years, so important for the Milanese astronomy. They were very important also for the culture and the society in general, since it was then that the Milanese Enlightenment flourished (see Table 1). A discussion of this movement and its relevance can be found for example in the recent book by Israel (2011). The periodical *Il Caffè*, i.e. Coffee House, was an important vehicle modelled after Joseph Addisons' *Spectator* and designed to promote the Enlightenment and its ideas. *Il Caffè* did not publish book reviews or news, however the Editor accepted to make an exception upon a request by Boscovich, defined in the periodical as "*uno de' più ragguardevoli Letterati d'Europa*" (one of the most respectable European scholars), to publish a review of the recent *Traité d'Astronomie* of Lalande (Verri 1766).

**Table 1.** Timeline of Milanese astronomy and Enlightenment.

| | |
|---|---|
| 1762 | Beginning of the astronomical research in Brera |
| 1763 | P. Verri, *Meditazioni sulla felicità* [Reflections on happiness] |
| 1763 | G. Parini, *Il Mattino (Il Giorno)* [The Morning (The Day)] |
| 1763 | Boscovich is appointed at the University of Pavia |
| 1764 | C. Beccaria, *Dei delitti e delle pene* [On Crimes and Punishement] |
| 1764 | Project of the Brera Observatory |
| 1765 | G. Parini, *Il Mezzogiorno (Il Giorno)* [The Midday (The Day)] |
| 1765 | Building of the Brera Observatory |
| 1764-1766 | *Il Caffè* by P. Verri et al. [*The Coffee House*] |

## 4. From 1765 to 1772

The *Risposta* written by Boscovich in 1772 (see next Section) contains detailed information on the observatory. At the upper level there was a large octagonal hall with six large windows. Some refracting and two reflecting telescopes were installed in the hall, and one telescope (Short) had an achromatic objective micrometer; moreover, such a large room contained clocks, instruments for optics experiments, celestial maps and models of instruments intended also for the public education, such as a wooden large parallactic machine. There was also a heliostat built by Boscovich both for performing experiments and for public observations of solar spots and eclipses. Above the flat roof there were two small towers that had a conical shape. A transit instrument, a parallactic machine and a clock were installed in one tower, while the large sextant of Canivet (with two telescopes) and a clock were installed in the other. Some slots on the flat roof allowed the observations with the telescopes placed in the hall. Four rooms were located downstairs under the hall: two were used as private rooms (with bed, chair, table and bookshelf) of observing astronomers, one contained meteorological instruments and under the other there was a small stock room. Stairs and balconies completed the building, along with two side rooms: one was essentially an entrance hall, and the other contained the large mural quadrant, a telescope and a clock. The research activity in this period included the determination of the latitude and longitude of the site and the related observations of the eclipses of Sun, Moon, Jupiter satellites and stellar occultations, some observations of comets (probably that of Lexell of 1770), Boscovich's method for the determination of comet orbits, and the problem of astronomical refractions. An intense effort concerned the development of methods and techniques for the determination of the instrument errors for a better reliability of the measurements. Meteorological observations were also performed regularly.

The education, previously a responsibility mainly of religious orders (and of the Senato of Milan as regards the University of Pavia), had been progressively taken in charge by the Austrian Government. The courses of applied mathematics were moved from Pavia to the Scuole Palatine in Milan in 1769, with a chair for Boscovich (optics, gnomonics and astronomy) and one for Frisi (hydraulics, engineering). Boscovich gave lessons and organized a laboratory of optics in Brera. An interesting testimony of such an activity is that of Charles Burney. He was a musician and historian of music, and also very interested in science; for example he wrote an essay for the history of comets. Burney was making a tour in France and Italy to collect information and documentation on the music in those countries, since he had planned to write a history of music. According to the experts in the field, his work was one the most important on that subject. During his stay in Milan, on july 1770 he visited the observatory, and his description of a working day in Brera is, as far as we know, unique in the literature. Burney attended at the experiments of optics that Boscovich was performing with the heliostat, and noted many young assistants, presumably students. He wrote an enthusiastic opinion: "If any new discoveries are to be made in astronomy, they may be expected from this learned Jesuit; whose attention to the optical experiments for the improvement of glasses, upon which so much depends; and whose great number of admirable instruments of all sort, joined to the excellence of the climate, and the wonderful sagacity he has discovered in the construction of his observatory and machines, form a concurrence of favourable circumstances, not easily to be found elsewhere" (Burney 1773). In 1769 Boscovich had been in Paris and then Bruxelles trying to solve the health problems regarding his leg. According to Burney (1773), "he was refused admission into the French academy, when at Paris, though a member, by the parliament, on account of his being a Jesuit; but if all Jesuits were like this father, making use only of superior learning and intellects for the avancement of science, and the happiness of mankind, one would have wished this society to be as durable as the world". These remarkable opinions could leave some doubts since Burney was not strictly a scientist. Actually they coincide with those of other scholars and astronomers such as Lalande (Proverbio 1987), and those of Schiaparelli (1938), who a century later wrote an essay (published posthumously) on the activity of Boscovich in Milan. "*Così in breve* *tempo era sorto in Milano un Osservatorio, per quell'epoca assai ben costituito. E' certo che se il Boscovich avesse potuto condurre a compimento tutte le sue idee, e se egli fosse stato lealmente e vigorosamente secondato, l'Osservatorio di Milano avrebbe potuto fin da quel tempo essere uno dei primi, o forse il primo, almeno sul Continente*". An observatory was made in Milan in a very short time, and it was very well planned. Surely, if Boscovich had been able to realize all his ideas, and if he had been strongly supported, the observatory of Milan would have been one of the first, or may be even the first, at least in the Continent. Schiaparelli then added "*ma le passioni umane entrarono in mezzo ad impedire un sì bell'esito*"; however, the human passions prevented such a beautiful outcome.

**5. Leaving from Milan**

There was an increasing disagreement between Boscovich and other Jesuits of Brera, such as Lagrange, and part of the difficulties probably derived from Boscovich's character. While the chair at the Scuole Palatine allowed him to stay close to the observatory (actually he probably lived in Brera), that may have exacerbated the problems related to the contrasting opinions on its management. Several complaints arrived in Vienna, where the Government was beginning to take care of the probable future management of the large Brera Collegium, given the difficulties of Jesuits. In 1771 the chancellor Kaunitz wrote to his minister plenipotentiary in Milan, Firmian, a letter where objections were risen concerning Brera (Proverbio 1987). He complained the few published works, the low quality of the astronomical measurements, and the use of the observatory just for a sterile show ("*sterile spettacolo*"). He requested a new plan, that is, Boscovich had to prepare what could be called a development plan. We may presume that Kaunitz would have liked to satisfy the needs of a serious scientific research in an observatory that had to be of public utility and not just a private institute. For example the astronomers should have made the observations needed for cartography (Proverbio 1987), and this will occur fifteen years later for the map of the Duchy. His criticism therefore had to be intended in a constructive sense, but Boscovich considered it a personal offence. He sent to Firmian the requested plan as a part of a very detailed document, the *Risposta* to a paragraph in a letter by Prince Kaunitz, which is very useful for the history of the observatory. It gives information on the research activity and its organization, it includes a description of the Specola and the astronomical instrumentation. The plan for the future includes the recruitment of new personnel, the tests still to be performed on the available instrumentation, the new instruments needed, the request for an adequate library, the observing programs. Here we comment briefly two points: the problems with the quadrant, and the activity on the public outreach.

The description of the painful difficulties with the quadrant of Canivet is surprising: "*Per le divisioni del nostro quadrante ... ho faticato infinitamente, nè fin ora ho potuto avere alcuna sicurezza de' metodi adoperati*". In spite of the efforts, Boscovich was not sure about the accuracy obtained[4]. Several problems with such an instrument derived from its structural flexures. The astronomers spent some years devising many methods and tools for the calibration, the accurate division in angular units, and the error analysis. Boscovich wrote several study reports describing the techniques devised by him.

To Kaunitz's objection of the observatory as "*oggetto di puro sterile spettacolo*", Boscovich replied that he designed the Specola just also for its use as a show, but he never considered it sterile, as his purpose was to give idea to the people about the astronomical instruments and observations: "*Io veramente nel darne il disegno ho avuto espressamente in vista che potesse anche servire di spettacolo; ma non l'ho mai creduto sterile ... il fine mio si era di dar idea nel paese degli istrumenti, dell'uso loro, e delle osservazioni astronomiche ... vi ho fatta fare una gran sala ottangolare, appunto, per potervi far dentro delle osservazioni in presenza di un numero considerabile di*

---

[4] "*Vi è ancora da faticar molto, e spender molto, prima che sia ridotto all'ultima perfezione ... Se l'avessi voluto adoprare senza queste precauzioni, avrei dato un gran numero di osservazioni erronee, o dubiose, come fanno tanti altri, che si fidano degli Artefici non mai totalmente esatti nei loro lavori*" (Proverbio 1987, p. 216).

*spettatori*". The large octagonal room was designed by Boscovich indeed to allow the participation of many visitors at the observations. He concluded recalling his indefatigable activity on the public outreach: "*Quanti Signori, quanti Religiosi non ho mai serviti di persona la su spiegando ogni cosa ... Questa è stata fra le tante mie occupazioni e cure una delle più assidue e faticose*". The public outreach and education was continued after Boscovich, but less intensively. During the first half of 19th century a public school of astronomy was operative, and during the twenties of the last century a Planetarium was built in Milan. However, it is only from about twenty years that specific resources, buildings and personnel in the observatory are dedicated to the public outreach, with the same spirit of Boscovich's time.

Boscovich was very upset, and in spite of the attempts of Firmian for a reasonable solution following Kaunitz's decisions concerning the direction of the observatory, in 1772 he gave up the chair and his duties. Some months later the Jesuit Order was suppressed by the Pope, and the friends in France were able to offer to Boscovich a prestigious assignment in Paris, director of naval optics of French marine. Many of the proposals of Boscovich regarding the observatory of Brera were accepted by Kaunitz, and it must be emphasized the thoughtfulness and care of Vienna for the development of the Specola, until the Napoleonic wars. For example, the optics for an achromatic telescope were purchased in England, and the telescope mounting was realized by Giuseppe Megele, an engineer who was moved from Vienna to Milan. Also an equatorial telescope of Sisson was purchased, and it was used (with some improvements) for almost a century; with this instrument Schiaparelli discovered Esperia in 1861. When Lagrange retired in 1776, the astronomers working in the observatory were his collaborator Francesco Reggio, Boscovich's pupil Angelo De Cesaris, and the young Barnaba Oriani, a Barnabite priest that had been Frisi's pupil. For an historical outline of the observatory in 18th and 19th centuries, see Antonello (2010; in italian).

In France, Boscovich studied in particular the achromatic telescopes, and he kept in touch with the astronomers of Brera. He prepared his monumental work *Opera pertinentia*, in which his new and old studies in optics and astronomy were collected. In 1782 he got the permission for a two years' leave from France for printing this work in Italy; then he got other extensions, and he went no more back to France.

## 6. The last years

After having spent two years in Tuscany, in the Papal Sates and the Venetian Republic, in 1785 he came back to Milan. Boscovich wished to complete the commentary to the philosophy of Benedetto Stay and he lived near Brera for a more convenient consultation of the books in the Braidense Library. He had begun the commentary many years before and it was published only in part. He considered this as his most important work. However, the health worsened and he suffered also of increasing mental disorders. After a short period in a mental hospital, Boscovich died in 1787, and was buried in Santa Maria Podone; unfortunately, there are no more indications of his grave. According to Schiaparelli (1938), in the lucid moments of the last months he regretted having spent his life time studying instead of devoting himself to spiritual matters. One could wonder if he was really lucid, or whether the letters of the last years contained already some clues of a repentance.

In her work, Hill (1961) complained of the non availability of Boscovich' letters, concluding that her biographical essay was necessarily incomplete and psychologically Boscovich remained largely unrevealed. For example, according to Hill, only his enemies and smaller minds, perhaps envious of his success, described him as vainglorious. Schiaparelli (1938) however concluded his work quoting Camillo Ugoni, who had remarked that vainglory was actually the main defect of Boscovich. It seems that Hill, who wrote a good and detailed biography, did not read, or had not the possibility of reading the documentation regarding Boscovich preserved in the Brera Observatory, and in particular the work of Schiaparelli (1938) that was based on several letters. According to Ugoni, Boscovich had a very good health ("*una salute erculea*"), but this does not appear to be true.

He wrote: "*La passione che lo invase per tutta la vita fu una stemperata ansietà di gloria ... A questa idolatria di gloria accoppiò egli una grande vivacità, onde i suoi colloqui erano tutto fuoco. ... Di sognate offese eragli fabbro la fantasia, al soperchio ardor della quale vuolsi ascrivere tale difetto ... Il massimo [dei suoi difetti] fu la vanagloria; e traspare da tutte le parole, da tutti gli scritti, da tutta la vita di lui: però che parlando, parlava sempre di sé, e sempre lodandosi, e si argomentava di provare ben anche alle dame quale Geometra egli fosse ... Riscattò il Boscovich ampiamente questo difetto con doti bellissime; pur tuttavia gli nocque assai, e più in Francia perché la lode è tributo che il mondo paga a grande stento, se lo esigono i creditori*". During all his life he was obsessed by an unrestrained passion for glory. To this idolatry he added a big liveliness so that his discussions were heated. His fantasy created unreal suffered offences, a defect that usually is ascribed to excessive imagination. His major defect was the vainglory, and all his words, all his writings, all his life betray it. He spoke always about himself, and always boasting, and trying to show even to the ladies how an important mathematician he was. His redeeming features were other very beautiful gifts; however such a defect damaged his reputation, especially in France, since the praise is a tribute that the world renders with difficulty if it is claimed by the creditor.

As a conclusion, let me quote Hill (1961) once again. She hoped that the unpublished manuscripts and large number of letters (that were still in private hands) would have belonged to some public institution: would it not be desirable in the interests of the history of science that a copy of all his works be deposited in important museums and libraries? She remarked that the letters of this exceptional eighteenth-century scholar with so wide a circle of correspondents among great figures that are part of the history of Europe should be published, and that would have been a service to scholarship and a worthy international monument to Boscovich.

It seems to me that now scholars are responding to such a plea. The National Edition (Edizione Nazionale delle Opere e della Corrispondenza di R.G. Boscovich[5]) is collecting and publishing the correspondence, the manuscrpits, the works. The publications are in italian, and it would be important to translate them in other languages.

---

[5] http://www.edizionenazionaleboscovich.it/